\documentclass [11pt]{article}
\usepackage{amsmath}
\usepackage{amsfonts}
\usepackage{amssymb}
\usepackage{graphicx}
\makeatletter \oddsidemargin  0in \evensidemargin 0in
\newcommand{\singlespacing}{\let\CS=\@currsize\renewcommand{\baselinestretch}{1.5}\tiny\CS}
\makeatletter \oddsidemargin  -.1in \evensidemargin -.1in
\newcommand{\doublespacing}{\let\CS=\@currsize\renewcommand{\baselinestretch}{1.35}\tiny\CS}

\def\@citex[#1]#2{\if@filesw\immediate\write\@auxout{\string\citation{#2}}\fi
  \def\@citea{}\@cite{\@for\@citeb:=#2\do
    {\@citea\def\@citea{,\linebreak[0]\hskip0pt plus .2em}%
      \@ifundefined{b@\@citeb}%
    {{\bf ?}\@warning{Citation `\@citeb' on page \thepage\space undefined}}%
      \hbox{\csname b@\@citeb\endcsname}}}{#1}}

\newtheorem{rule-def}[theorem]{Rule} 
\begin{document}
\title{\bf A cloned qutrit and its utility in information processing tasks}
\author{Sovik Roy $^{1, 3}$\thanks{sovikr@rediffmail.com},\:\:\: Nirman Ganguly $^{2, 3}$\thanks{nirmanganguly@gmail.com},\\\\ Atul Kumar$^{4}$\thanks{atulk@iitj.ac.in}\\\\ Satyabrata Adhikari $^{4}$\thanks{satya@iitj.ac.in},\:\:\:  A. S. Majumdar $^{3}$\thanks{archan@bose.res.in}\\\\
$^1$ Techno India, Salt Lake City, Kolkata - 91, India \\
$^2$ Heritage Institute of Technology, Kolkata - 107, India\\
$^3$ S. N. Bose National Centre for Basic Sciences, Salt Lake,
Kolkata-98, India\\
$^4$ Indian Institute of Technology, Rajasthan,
Jodhpur- 11, India\\
}
\maketitle
\begin{abstract} 
\noindent We, in this paper, analyze the efficacy of an output as a resource from a universal quantum cloning
machine in information processing tasks such as teleportation and dense coding. For this,
we have considered the $3\otimes 3$ dimensional system (or qutrit system). The output states are found to be NPT states for certain ranges of machine parameters. Using the output state as an entangled
resource, we have also studied the optimal fidelities of teleportation and capacities of dense coding
protocols with respect to the machine parameters and have made a few interesting observations.
Our work is mainly motivated from the fact that the cloning output can be used as a resource in
quantum information processing and adds a valuable dimension to the applications of cloning
machines.
\end{abstract}
\section{Introduction}
Quantum entanglement, apart from being central to the investigations of the fundamentals of quantum mechanics, has also been used as a resource, enabling efficient quantum information processing protocols such as quantum teleportation \cite{Bennettteleport}, superdense coding \cite{BennettDensecoding,mattle761996}, cryptography \cite{BennettCrypto,ekert671991}, quantum cloning \cite{Wootterscloning} etc. The above protocols were all developed using qubits as fundamental units of quantum information.The question of usefulness of states for quantum teleportation \cite{Lindenteleport,horodecki601999}, dense coding \cite{Liudensecoding} and quantum cryptography \cite{bruss2003,acin2005} have also been studied in details. Over the last few years, a growing interest has been felt in quantum information community to study  multi-level or higher dimensional systems or continuous spectrum systems for doing quantum information processing. For example the concept of quantum cloning has been extended from qubits to qutrits (quantum three level systems) \cite{Cerfcloning}. This is because, in the higher dimension, quantum information processes are supposed to be more efficient in certain situations. A quantum state in a large dimensional space contains more information than one in small dimensional space \cite{Wiseman}. The present experimental context makes it reasonable to consider the manipulation of more-than-two-level quantum information carriers.\\\\
Keeping the recent trends in mind, we, in this article, have focussed on developing a state in qutrit system, whose capability in information processing has been studied. Central to our investigations, is the usefulness of mixed states of two qutrits, obtained as an output from Buzek - Hillery universal quantum cloning machine \cite{buzek221998}, as resources for quantum teleportation and dense coding protocols. Our analysis of teleportation and dense coding using higher dimensional systems provides some interesting aspects about the outputs obtained from the cloning machine when used as resources in quantum information processing. We show that the non-optimal output state obtained from the Buzek - Hillery cloning machine can be used more efficiently as a resource for quantum information processing in comparison to the optimal output state obtained from the Buzek - Hillery cloning machine. Surprisingly, our results show that the optimal teleportation fdelity for a distilled non-optimal output state as a resource is more than the optimal teleportation fdelity for a distilled optimal output state as a resource for a certain range of machine parameters. For dense coding one can successfully use only a distilled non-optimal output state as a resource and not the distilled optimal output state. We believe that the results obtained in this article would add another important dimension to the applicability of quantum cloning machines to quantum information processing \cite{adhikari2008}.\\\\
In section $2$ we give a brief review of Buzek-Hillery universal quantum cloning machine and discuss the entanglement properties of the output states obtained from the cloning machine. Sections $3$ and $4$ deal with the analysis of the optimal and non-optimal output state, respectively, for quantum information processing. We conclude the article with results and discussions in section $5$.
\section{A brief review of Buzek - Hillery universal quantum cloning machine (BH-UQCM):}
 To begin with, we consider the universal quantum cloning machine (UQCM) for arbitrary higher dimensional Hilbert space suggested by V. Buzek and M. Hillery \cite{buzek221998}. The BH-UQCM is  an $n$ - dimensional quantum system, where $\vert X_{i}\rangle_{x}, (i = 1,2, \ldots ,n)$ is an orthonormal basis of the cloning machine Hilbert Space. If we consider that the cloner is initially prepared in a particular state $\vert X\rangle_{x}$, then the transformation  for the basis vectors corresponding to the B-H cloning machine is given by\\
\begin{eqnarray}
\vert\psi_{i}\rangle_{a}\:\vert0\rangle_{b}\:\vert X\rangle_{x}\longrightarrow c\:\vert\psi_{i}\rangle_{a}\:\vert\psi_{i}\rangle_{b}\:\vert X_{i}\rangle_{x} + d\:\sum_{j\neq i}^{n}\:\lbrace\vert\psi_{i}\rangle_{a}\:\vert\psi_{j}\rangle_{b} + \vert\psi_{j}\rangle_{a}\:\vert\psi_{i}\rangle_{b}\:\rbrace\:\vert X_{j}\rangle_{x}\label{BHT},
\end{eqnarray} \\
with real coefficients $c$ and $d$. The action of the cloning transformation on a state can be specified by a unitary transformation acting on the basis vectors of the tensor product space of the original quantum system $\vert\psi_{i}\rangle_{a}$, the copier and an additional $n$ - dimensional system which becomes the copy (which is initially prepared in a specific state $\vert0\rangle_{b}$). From the unitarity of the above transformation it follows that $c$ and $d$ satisfy the relation\\
\begin{eqnarray}
c^{2} + 2\:(n-1)\:d^{2} = 1 \label{cdrel1}.
\end{eqnarray}\\
Using the transformation it was shown that the particles $a$ and $b$ at the output of the cloner are in the same state (having the same reduced density matrices), which was described by the density operator\\
\begin{eqnarray}
\widehat{\rho}_{out}^{a} = \widehat{\rho}_{out}^{b} =\sum_{i=1}^{n}\vert\alpha_{i}\vert^{2}[c^2 + (n-2)d^2]\vert\psi_{i}\rangle\langle\psi_{i}\vert + \sum_{i,j=1,i\neq j}^{n}\alpha_{i}\alpha_{j}^{*}[2cd + (n-2)d^2]\vert\psi_{i}\rangle\langle\psi_{j}\vert + d^2I\label{genoutput}.
\end{eqnarray}\\
For universal cloning transformation which generates two imperfect copies from the original state (say, $\vert\phi\rangle_{a}$), the quality of the cloning will not depend on the particular state (in the given Hilbert space) which is going to be copied if and only if the output reduced density matrix is of the form\\
\begin{equation}
\widehat{\rho}_{out}^{j} = s\:\widehat{\rho}_{in}^{j} + \frac{1-s}{n}\:I \label{scaling form},
\end{equation}\\
where $\widehat{\rho}_{in}^{j} = \vert\phi\rangle\langle\phi\vert$ is the density operator describing the original state which is going to be cloned. Here the quantity $s$ is called the scaling factor. To find the values for the parameters $c$ and $d$ the density operator in equation (\ref{genoutput}) must take the scaled form of (\ref{scaling form}). This directly guarantees the universality of the transformation (\ref{BHT}). Comparing equations (\ref{scaling form}) and (\ref{genoutput}) it has been found in \cite{buzek221998} that the parameters $c$ and $d$ satisfy the equation\\
\begin{equation}
c^{2} = 2\:c\:d \label{cdrel}.
\end{equation} \\
Then the normalization condition in equation (\ref{cdrel1}) gives the explicit values for the parameters $c$ and $d$ which are given as follows:\\
\begin{equation}
c^2 = \frac{2}{n+1},~~~~~~~d^2 = \frac{1}{2(n + 1)} \label{cdrel2},
\end{equation}\\
from which it follows also that the scaling factor $s$ is given by\\
\begin{equation}
s = c^2 + (n - 2) d^2 = \frac{n + 2}{2 (n + 1)}\label{scaling form 1}.
\end{equation}\\
Equation (\ref{cdrel2}) gives the optimal values for the real parameters $c$ and $d$. For qutrit system we have $n = 3$ in equation (\ref{cdrel1}) and hence $c^2 = \frac{1}{2}$ and $d^2 = \frac{1}{8}$ (from equation (\ref{cdrel2})).
\section{Analysis of two qutrit output from BH - UQCM :}
We now proceed to discuss the entanglement properties of a two qutrit system generated as the output of the UQCM (\ref{BHT}). We consider a single qutrit system as an input to the cloning machine, namely\\
\begin{eqnarray}
\vert\varphi\rangle = \frac{1}{\sqrt{3}}\:\lbrace\:\vert0\rangle + \vert1\rangle + \vert2\rangle\:\rbrace. \label{singlequtrit}
 \end{eqnarray}\\
The basis vectors $\vert0\rangle$, $\vert1\rangle$ and $\vert2\rangle$ have their independent transformed forms with respect to (\ref{BHT}) whereas from the linearity property of the cloning transformation and then on tracing out the machine vectors $X_{i}, i = 1,2$, as well as  by considering only the unitarity condition (\ref{cdrel1}) on $c$ and $d$ for $n = 3$ we get the composite two qutrit density operator as\\
\begin{eqnarray}
\rho_{ab}^{out} = \frac{(1-4\:d^{2})}{3}\:\sum_{i=0}^{2}\:\vert i,i\rangle\langle i,i\vert + \frac{2}{3}\:d^{2}\:\sum_{i\:\neq\:j}^{2}\:[\:\langle i,j\vert + \langle j,i\vert\:] \nonumber\\
+\frac{(\sqrt{1-4\:d^{2}})}{3}\: d\:[\sum_{i\: \neq \: j}^{2}\:\vert i,i\rangle\:(\langle i,j\vert\: + \langle j, i\vert\:)+\sum_{i\:\neq\:j}^{2}\:\vert i,j\rangle\:(\langle i,i\vert + \langle j,i\vert)]\nonumber\\ + \frac{d^{2}}{3}\:[\sum_{i\:\neq\:j}^{2}\:\vert i,j\rangle\:(\:\langle i,j+1\vert + \langle j+1,i\vert + \langle i+1,j+1\vert +\langle j+1, i+1\vert\:)\:(mod\:2)\:].\label{rhoout}
\end{eqnarray}\\
The above state is a function of machine parameter $d$.\\\\ Now by positive partial transposition criteria \cite{peres1996}, with respect to the system $a$, we find that there exists two eigenvalues of the state (\ref{rhoout}) as\\
\begin{eqnarray}
e_{1} &=& \frac{1+4d^{2}}{6}-\frac{1}{6}\sqrt{1+24d^{2}-104d^{4}+32\sqrt{-(2d-1)(2d+1)}d^{3}} \nonumber\\ \nonumber\\
e_{2} &=& \frac{1-5d^{2}}{6}-\frac{1}{6}\sqrt{1-6d^{2}+25d^{4}-16\sqrt{-(2d-1)(2d+1)}d^{3}}.
\label{e1e2}
\end{eqnarray}\\
It is observed that at least one of the two obtained eigenvalues are always negative when $d\:\in\:(0,\:\frac{1}{2}]$. The eigenvalue $e_{1}$ is negative when $d \in (0, \frac{6+\sqrt{2}}{17})$ and $e_{2}$ is negative when either $d \in (0, \frac{1}{2\sqrt{2}})$ or $d \in (\frac{1}{2\sqrt{2}}, \frac{1}{2}]$. Therefore the state (\ref{rhoout}) is NPT state for $d$ belonging to either of these ranges. Combining the above results we can say that the non - optimal output (\ref{rhoout}) is an NPT state for $d \in (0, \frac{1}{2}]$. The fact that the cloned two-qutrit system generated from the UQCM is a NPT state motivates us to analyse its utility for quantum information processing tasks such as teleportation and dense coding.\\\\
Also we know that if $\rho^{1}$ represents the ideal density operator describing the \textit{in} state which will pass through a certain cloning machine and $\rho^{2}$ is the density operator of the output state from that machine, then fidelity of cloning is calculated by \textit{Bures' distance} \cite{bures1969} and is defined as follows\\
\begin{eqnarray}
d_{B}(\rho^{1},\rho^{2})=\sqrt{2}\:\lbrace 1-Tr\:\sqrt{(\rho^{1})^{\frac{1}{2}}\:\rho^{2}\:(\rho^{1})^{\frac{1}{2}}}\:\rbrace^{\frac{1}{2}} \label{bures1}.
\end{eqnarray}\\
So using (\ref{bures1}), the fidelity of cloning of the non - optimal two qutrit output $\rho_{out}^{ab}$ of (\ref{rhoout}) , is then calculated as\\
\begin{equation}
d_{B}(\rho_{out}^{a},\rho_{in}^{a}) = \frac{\sqrt{2}}{3}\:\sqrt{\:9-3\:\sqrt{\:3 + 6\:d^2 +12\:d \:\sqrt{\:1-4\:d^2}}} \label{bures2},
\end{equation}\\
where $\rho_{in}^{a} = \vert \varphi\rangle \langle \varphi \vert$ and $\rho_{out}^{a}$ is the reduced density operator of $\rho_{out}^{ab}$ which is obtained by tracing out party $b$. We see that fidelity of cloning of state (\ref{rhoout}) is also a function of machine parameter $d$. BH - UQCM is such a quantum copier where\\
\begin{equation}
d_{B}(\rho_{out}^{a},\rho_{in}^{a}) = constant \label{bures3}.
\end{equation}\\
This also implies that for different values of the machine parameter $d$ we get different forms of the non - optimal output (\ref{rhoout}) and hence different fidelities of cloning. 
From equation (\ref{bures2}), we see that, when $d^{2} = \frac{1}{8}$, the fidelity of cloning from is given by $0.517638$ and which is similar to the value obtained after substituting $n=3$, in the following expression.\\
\begin{equation}
d_{B}(\rho_{out}^{a},\rho_{in}^{a}) = \sqrt{2} \sqrt{1 - \sqrt{\frac{n + 3}{2 (n + 1)}}}\label{bures4}
\end{equation}
\subsection{Quantum teleportation and dense coding using the optimal output state of BH-UQCM:}
The output two-qutrit state obtained from the cloning machine will be an optimal state (which we denote by $\rho^{ab}_{optout}$)  for $d^{2} = \frac{1}{8}$. From (\ref{rhoout}), it is easy to show that such an output state would be an NPT state according to \cite{peres1996}.\\\\
In n-dimensional system, a state $\rho$ can be used as a teleportation channel, if its fully entangled fraction i.e. $F(\rho)$ is greater than $\frac{1}{n}$ \cite{horodecki591999}. The fully entangled fraction for any arbitrary state $\rho$ is defined as\\
\begin{equation} F(\rho)=Max_{\phi}\langle\phi\vert\rho\vert\phi\rangle \label{fef},
\end{equation} \\
where maximum is taken over all the maximally entangled basis states $\phi$. In this respect our two qutrit output state is supposed to be useful in teleportation if its fully entangled fraction is more than $\frac{1}{3}$. To check the utility of the two qutrit optimal output state $\rho_{optout}^{ab}$ in  teleportation , we therefore calculate its fully entangled fraction.\\\\
We know that the Bell basis states are four orthonormal maximally entangled states of two qubits, which are given by\\
\begin{equation}
\vert\phi^{\pm}\rangle =\frac{1}{\sqrt{2}}\lbrace \vert0,0\rangle \pm \vert1,1\rangle\rbrace \label{Bell states 1},
\end{equation}\\
and
\begin{equation}
\vert\psi^{\pm}\rangle =\frac{1}{\sqrt{2}}\lbrace \vert0,1\rangle \pm \vert1,0\rangle\rbrace \label{Bell states 2}.
\end{equation}\\
Analogous to these above states, for $3-$ dimensional systems, or qunits we have \cite{karimipour2006}\\
\begin{equation}
\vert\phi_{xy}\rangle =\frac{1}{\sqrt{3}}\:\sum_{j=0}^{2}\:\xi^{jy}\vert j,j+x \rangle,~~~~ x, y =0,\:1,\:2 \label{ndimbasis},
\end{equation}\\
where, $\xi := e^{\frac{2\pi i}{3}}$ and $\lbrace \vert 0 \rangle, \vert 1 \rangle, \vert 2\rangle\rbrace$ is an orthonormal basis for the space of one qutrit. The states (\ref{ndimbasis}) are maximally entangled and are mutually orthogonal.\\\\
Hence considering $\rho_{optout}^{ab}$ and using (\ref{fef})  and (\ref{ndimbasis}) and using the formula\\
\begin{eqnarray}
\langle\phi_{ij}\vert\rho\vert\phi_{ij}\rangle = Tr(\rho\vert\phi_{ij}\rangle\langle\phi_{ij}\vert). \label{formula1}
\end{eqnarray} \\
we calculate all the nine inner products of optimal output. We find that,  $F(\rho_{optout}^{ab}) = \frac{1}{6} < \frac{1}{3}$, implying that the state $\rho_{optout}^{ab}$  is not useful for quantum teleportation teleportation tasks.\\\\
We now proceed to analyze whether the optimal state is useful for dense coding or not.
A state is said to be 'dense codeable' if it can be used in dense coding \cite{bruss212004}. In $n_{a}\otimes n_{b}$ systems, the capacity of dense coding for any given shared state $\rho^{ab}$ has been defined as, \\
\begin{eqnarray}
\chi = \log_{2}n + S_{V}(\rho^{b}) - S_{V}(\rho^{ab}) \label{densecodeability}.
\end{eqnarray}\\
The term $S_{V}(\rho)$ denotes the von - Neumann Entropy, where $S_{V}(\rho^{b})$ is the von - Neumann entropy of the reduced system and $S_{V}(\rho^{ab})$ is von - Neumann entropy for the original one (i.e. the joint state $\rho^{ab}$). von - Neumann Entropy is considered to be the standard measure of randomness of a statistical ensemble described by a density matrix. For any arbitrary state $\rho$ this is denoted by calculated as,\\
\begin{eqnarray}
S_{V}\:(\rho) = -tr\:(\rho\:\log\:\rho) = -\sum_{i}\:k_{i}\:\log\:(k_{i}),
\label{von1}
\end{eqnarray}\\
where $k_{i}$' s are the eigenvalues of the state $\rho$ along-with the condition that $0\equiv 0 \log 0$.  This measure has a natural significance stemming from its connections with statistical physics and information theory.\\\\
 A shared quantum state is thus said to be useful for dense coding , if the corresponding capacity $\chi$ is more than $log_{2}\:(n)$ (in qutrit system it is $log_{2}\:(3)$) . From (\ref{densecodeability}), it is clear that such states are precisely those for which $S_{V}(\rho^{b}) - S_{V}(\rho^{ab}) > 0$.
For the state $\rho_{optout}^{ab}$, we find that $S_{V}(\rho_{optout}^{b})-S_{V}(\rho_{optout}^{ab}) = -0.43872 < 0$. Hence the state $\rho_{optout}^{ab}$ is not useful for the dense coding protocol. As the optimal output state is neither useful for teleportation nor useful in dense coding, we proceed further for the distillation of the optimal output state and to investigate its efficacy for teleportation as well as for dense coding.\\
\subsection{Construction of filter and analysis of the distilled optimal  state in teleportation and dense coding:}
In Section $3.1$, we have seen that the optimal output state $\rho_{optout}^{ab}$ is not useful for teleportation and dense coding protocols. For $2\otimes 2$ dimensional systems, any state can be made useful
for teleportation \cite{verstraete902003}.\\\\
Horodecki et al.\ \cite{horodecki591999} showed that any state violating the reduction criteria is distillable. In  reduction criteria, for a state $\rho$, we first calculate the state $\rho^{a} = Tr_{b}(\rho)$, which is a reduction of the state of interest. Then one should check the non negativity of the eigenvalues of the operator $(\rho^{a} \otimes I - \rho)$ i.e. one should have
\begin{eqnarray}
\rho^{a} \otimes I - \rho \geq 0 \label{redcrit1}.
\end{eqnarray}
The dual criteria
\begin{eqnarray}
I \otimes \rho^{b} - \rho \geq 0 \label{redcrit2},
\end{eqnarray}
can also be used.  It is easy to see that the state $\rho^{ab}_{optout}$ violates reduction criteria. Hence it is distillable. We can now distill $\rho^{ab}_{optout}$ calculating the eigenvector corresponding to the suitable negative eigenvalue of the state, $(\rho^{a}_{optout} \otimes I - \rho^{ab}_{optout})$ and subjecting the state to the appropriate filter. For this, we need to calculate the eigen vector $\vert \Psi\rangle$ corresponding to the negative eigenvalue of the operator $(\rho^{a}_{optout} \otimes I - \rho^{ab}_{optout})$. If the form of such an eigen vector is $\vert \Psi\rangle=\sum_{i,\:j}^{N}\:a_{ij}\vert i\rangle\vert j\rangle$, then the filter $A$ is nothing but an operator which can simply be represented using a matrix where the element of the matrix can be given as $A_{ij}=\sqrt{N}\:a_{ij}$ \cite{horodecki591999}.  Also, we know that if $\rho$ is any state and A is a filter then a new state $\rho^{/}$ is found as\\
\begin{eqnarray}
\rho^{/}= \frac{A^{\dagger} \:\otimes\: I \:\rho A\: \otimes\: I}{Tr (\rho\: AA^{\dagger} \:\otimes \:I)} \label{filteredform},
\end{eqnarray}\\
where $A$ is the filter.
By following the procedures described in \cite{horodecki591999}, we construct the filter for the optimal state $\rho^{ab}_{optout}$. The filter is denoted by $A_{opt}$ and is given by\\
\begin{eqnarray}
A_{opt} = \left(%
\begin{array}{ccc}
 \sqrt{3}(\frac{3}{2}-\frac{\sqrt{29}}{2}) &  \sqrt{3}(-\frac{7}{2}+\frac{\sqrt{29}}{2}) &  -\sqrt{3}\\
 \sqrt{3}(\frac{7}{2}-\frac{\sqrt{29}}{2}) &  \sqrt{3}(-\frac{3}{2}+\frac{\sqrt{29}}{2}) &  \sqrt{3}\\
 \sqrt{3}(\frac{5}{2}-\frac{\sqrt{29}}{2}) &  \sqrt{3}(-\frac{5}{2}+\frac{\sqrt{29}}{2}) &  0\\
\end{array}%
\right)\label{optfilter}.
\end{eqnarray}\\
Using (\ref{filteredform}) we find the distilled form of the optimal state of $\rho^{ab}_{optout}$. Let us denote this  by $\varrho^{ab}_{optout}$ and so we have\\
\begin{eqnarray}
\varrho^{ab}_{optout} = \frac{A_{opt}^{\dagger}\: \otimes\: I\: \rho^{ab}_{optout}\: A_{opt} \:\otimes \:I}{Tr (\rho^{ab}_{optout} \:
A_{opt}\: A_{opt}^{\dagger}\: \otimes \:I)} \label{filteredoptrhoout}.
\end{eqnarray}\\
The filtered optimal state (\ref{filteredoptrhoout}) is found to be suitable for teleportation since the fully entangled fraction  of  $\varrho^{ab}_{optout}$ i.e. $ F(\varrho^{ab}_{optout}) = 0.38789 > \frac{1}{3}$. Moreover, the optimal teleportation fidelity for any arbitrary state $\rho$ in $n - $ dimensional system is defined as\\
\begin{eqnarray}
f(\rho) = \frac{n F(\rho) + 1}{n + 1} \label{telefidy}.
\end{eqnarray}\\
Therefore in qutrit system, the optimal teleportation fidelity of the state $\varrho^{ab}_{optout}$ is given by\\
\begin{eqnarray}
f(\varrho^{ab}_{optout}) =  \frac{3 F(\varrho^{ab}_{optout}) + 1}{4}=0.5409 \label{valueopttelefidy}.
\end{eqnarray}\\
Similarly, for the dense coding protocol using (\ref{densecodeability}) and (\ref{filteredoptrhoout}) we see that,  $S_{V}(\varrho^{b}_{optout}) - S_{V}(\varrho^{ab}_{optout}) = -0.3327 < 0$. Therefore the filtered optimal state is still not useful for dense coding.
\subsection{Quantum teleportation and dense coding using the non-optimal output state of BH-UQCM::}
For the values of machine parameter $d$ other than $\frac{1}{2\sqrt{2}}$ (i.e. $d^{2}= \frac{1}{8}$), the state (\ref{rhoout}) is said to be non - optimal. Also we know that for $d \in (0, \frac{1}{2}]$, the state (\ref{rhoout}) is NPT. Now the non optimal output state (\ref{rhoout}) cannot be used as a teleportation channel, for in this case we see that $F(\rho_{nonopt}^{ab})= \frac{4 d^{2}}{3}$ never exceeds $\frac{1}{3}$. Hence the non-optimal state $\rho_{nonopt}^{ab}$ is also not useful to be used as a resource for teleportation process. \\\\
This can be verified also in the following way. We know that a hermitian operator $W$ may be called a teleportation witness if the following conditions are satisfied (i) $Tr(W\sigma) \geq 0$, for all states $\sigma$ which are not useful for teleportation, (ii) $Tr(W\chi) < 0$ for at least one state $\chi$ which is useful for teleportation \cite{ganguly2011,satyawitness}, where\\
\begin{eqnarray}
W =  \frac{I}{3}-\vert \phi^{+}\rangle\langle \phi^{+}\vert \label{witness},
\end{eqnarray}\\
where, $\vert\phi^{+}\rangle = \frac{1}{3}\sum_{i=0}^{2}\vert ii \rangle$.\\\\
For the non-optimal state $\rho_{nonopt}^{ab}$, we see that $Tr(W\rho_{nonopt}^{ab})= \frac{4}{3} d^{2}$, which is always positive for $0 < d \leq \frac{1}{2}$. This proves that the state $\rho_{nonopt}^{ab}$ cannot be used in telelportation.\\\\
In order to see whether the state $\rho_{nonopt}^{ab}$ is useful in dense coding or not, we plot a graph between $S_{V}(\rho^{b}_{nonopt}) - S_{V}(\rho^{ab}_{nonopt})$ and the machine parameter $d$, where $\rho^{b}_{nonopt}$ is the reduced density operator of the two qutrit output state $\rho^{ab}_{nonopt}$ and is given by,\\
\begin{eqnarray}
\rho^{b}_{out} = \left(%
\begin{array}{ccc}
  \frac{1}{3} & \frac{1}{3}d(2\sqrt{1-4d^2}+d) & \frac{1}{3}d(2\sqrt{1-4d^2}+d) \\
  \frac{1}{3}d(2\sqrt{1-4d^2}+d) & \frac{1}{3} & \frac{1}{3}d(2\sqrt{1-4d^2}+d) \\
  \frac{1}{3}d(2\sqrt{1-4d^2}+d) & \frac{1}{3}d(2\sqrt{1-4d^2}+d) & \frac{1}{3} \\
\end{array}%
\right) \label{reducedrho}.
\end{eqnarray}\\
It is clear from the following figure that the non optimal two qutrit output state $\rho_{nonopt}^{ab}$ is not useful for dense coding when $d \in (0, \frac{1}{2}]$. 
\begin{figure}[!ht]
\centering
\resizebox{6cm}{6cm}{\includegraphics{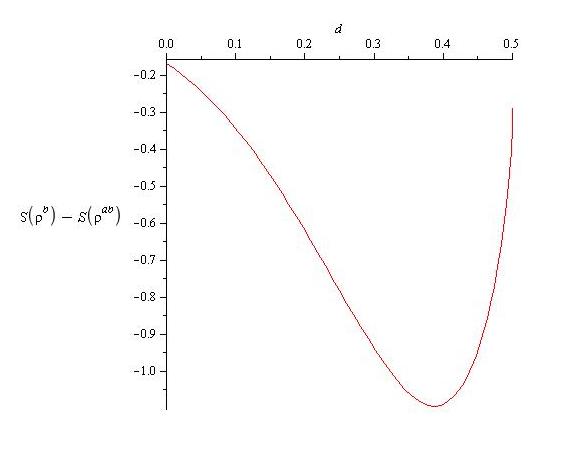}}
\caption{\footnotesize The figure shows that the difference $S_{V}(\rho^{b})-S_{V}(\rho^{ab})$ always lies in the fourth quadrant of the cartesian plane with respect to the values of $d$ in the range $(0, \frac{1}{2}]$.}
\end{figure}\newpage
\subsection{Construction of filter and study of the non - optimal filtered states in teleportation and dense coding:}
As the non-optimal output state also cannot be used for information transfer, we try to find out whether it is possible to distill the non-optimal output state using an appropriate filter so that the distilled system can be used for quantum information protocols.\\\\
Similar to the optimal output state, 
the non optimal state (\ref{rhoout}) violates the reduction criteria \cite{horodecki591999}. So as before we can  distill $\rho^{ab}_{nonopt}$ calculating the eigenvector corresponding to the suitable negative eigenvalue of the state, $(\rho^{a}_{nonopt} \otimes I - \rho^{ab}_{nonopt})$, subjecting the state to the appropriate filter. We find the eigenvalue as\\
\begin{eqnarray}
e_{1}^{/} = \frac{1-3d^{2}}{6}+
\frac{1}{3}\sqrt{-(2d-1)(2d+1)}d-\nonumber\\\frac{1}{6}\sqrt{1-18d^{2}+4\sqrt{-(2d-1)(2d+1)}d+113d^{4}-44d^{3}\sqrt{-(2d-1)(2d+1)}}
\label{eigen},
\end{eqnarray}\\
which is always negative for $d \in\:(\frac{6+\sqrt{2}}{17}, \frac{1}{2}]$. Hence following \cite{horodecki591999}, we construct the filter $A_{nonopt}$ to distill the state $\rho_{nonopt}^{ab}$ where,\\
\begin{eqnarray}
A_{nonopt} =  \sqrt{3}\left(%
\begin{array}{ccc}
 1 &  -k &  -k\\
-k &   1 &  -k\\
-k &  -k &   1 \\
\end{array}%
\right) \label{nonoptfilter1}.
\end{eqnarray}\\
where $k= \frac{11d^{2}-2\sqrt{1-4d^{2}}d + \sqrt{1- 18d^{2}+ 4\sqrt{1 - 4d^{2}}d+113d^{4}-44d^{3}\sqrt{1-4d^{2}}-1}}{4d^{2}}$ and $d\in\:(\frac{6+\sqrt{2}}{17},\:\frac{1}{2}]$.\\\\ This will transform the non optimal state $\rho^{ab}_{out}$ to its filtered form $\tau^{ab}_{out}$ defined by\\
\begin{eqnarray}
\tau^{ab}_{out} = \frac{A_{nonopt}^{\dagger}\: \otimes\: I\: \rho^{ab}_{out}\: A_{nonopt}\:\otimes\: I}{Tr (\rho^{ab}_{out} \: A_{nonopt}\: A_{nonopt}^{\dagger} \: \otimes I)} \label{filterednonoptrhoout1}.
\end{eqnarray}\\
The fully entangled fraction of $\tau^{ab}_{out}$ is given as\\
\begin{eqnarray}
F(\tau^{ab}_{out}) = \nonumber\\ \frac{4[d^{2}(2(1-t_{1})+d^{2}(22t_{1}-31)+t_{2}(10-110d^{2}-6t_{1})+198d^{4})]}{3[(1-k)+t_{2}(6-68d^{2}+94d^{4}-4t_{1}+12t_{1}d^{2})+d^{2}(6t_{1}-9-5d^{2}+23t_{1}d^{2})]} \label{feffilnonopt1},
\end{eqnarray}\\
where $t_{2}=\sqrt{1-4d^{2}}.d$ and $t_{1}=\sqrt{1-18d^{2}+4t_{2}+113d^{4}-44d^{2}.t_{2}}$\\\\
Evidently, for $\frac{6+\sqrt{2}}{17} < d \leq \frac{1}{2}$, the states $\tau^{ab}_{out}$ are always suitable for teleportation since there $F (\tau^{ab}_{out}) > \frac{1}{3}$. Moreover, the optimal teleportation fidelity of $\tau^{ab}_{out}$ is then given by\\
\begin{eqnarray}
f(\tau^{ab}_{out}) = \frac{3 F(\tau^{ab}_{out}) + 1}{4}  \label{telefidy1}.
\end{eqnarray}\\
Alternately, using \cite{ganguly2011,satyawitness} and (\ref{witness}) it has been observed that $Tr(W\tau^{ab}_{out})= -\frac{1}{4d^{4}}[-1+439d^{6}+17d^{2}-119d^{4}-534d^{4}t_{2}+65d^{4}t_{1}+108d^{2}t_{2}-6t_{2}+t_{1}+4t_{1}t_{2}-36d^{2}t_{1}t_{2}-14d^{2}t_{1}]$.\\\\
From there we find that $Tr(W\tau^{ab}_{out})< 0$ for $d \in (\frac{6+\sqrt{2}}{17}, \frac{1}{2}]$, which implies that in this range the state $\tau_{out}^{ab}$ can always be used in teleportation but for $d \in (0, \frac{6+\sqrt{2}}{17})$, the filtered state $\tau_{out}^{ab}$ cannot be used in teleportation since $Tr(W\tau_{out}^{ab}) > 0$ there.\\\\
In order to check the usefulness of the state $\tau_{out}^{b}$ for dense coding, we show that $S_{V}(\tau^{b}_{nonopt}) - S_{V}(\tau^{ab}_{nonopt})\:>\:0$ for $d\in \: (\frac{6+\sqrt{2}}{17}, \frac{1}{2}]$. Hence, the state $\tau_{out}^{ab}$ can also be used as a resource in dense coding protocol as also evident from the figure $3$. Interestingly, the optimal fidelity obtained in case of teleportation using the distilled non-optimal output state as an entangled resource is greater than the optimal fidelity obtained in case of teleportation using the distilled optimal output state. Moreover, the distilled optimal output state cannot be used for dense coding, but distilled non-optimal output state can be successfully used as a resource for dense coding.
\noindent
In the following figure we plot the teleportation fidelities and capacities of dense coding of non - optimal filtered states corresponding to $d$. We see that the fidelity of teleportation reaches its maximum for $d = \frac{1}{2}$ and the maximum value is $0.68$ approx whereas the for $d = \frac{1}{2}$ the capacity of dense coding of $\tau^{ab}_{out}$ is $2.08$ approx.\\
\begin{figure}[!ht]
\centering
\resizebox{8cm}{8cm}{\includegraphics{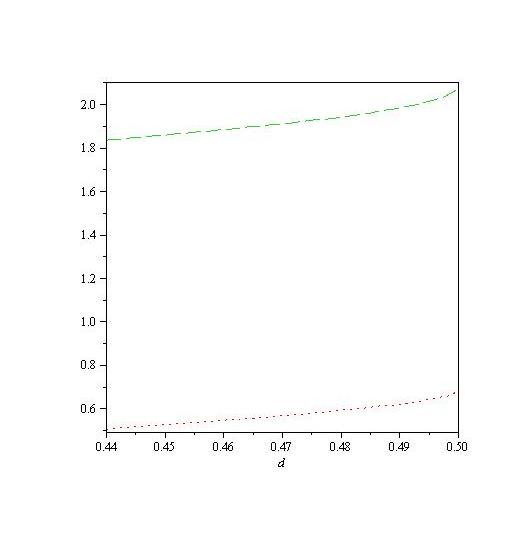}}
\caption{\footnotesize In the figure the dotted lines represent the capacity of dense coding i.e. $\chi(\tau^{ab}_{out})$ of the non optimal states and dashed line represent the teleportation fidelities i.e $f(\tau^{ab}_{out})$ of the said state with respect to $d$.}
\end{figure}\\
Now using (\ref{von1}) we calculate the mixedness of the filtered non optimal state $\tau_{out}^{ab}$ as before, and we plot the mixedness of $\rho_{out}^{ab}$ and $\tau_{out}^{ab}$ against the parameter $d \in (\frac{6+\sqrt{2}}{17}, \frac{1}{2}]$ in the following.
\begin{figure}[!ht]
\centering
\resizebox{6cm}{6cm}{\includegraphics{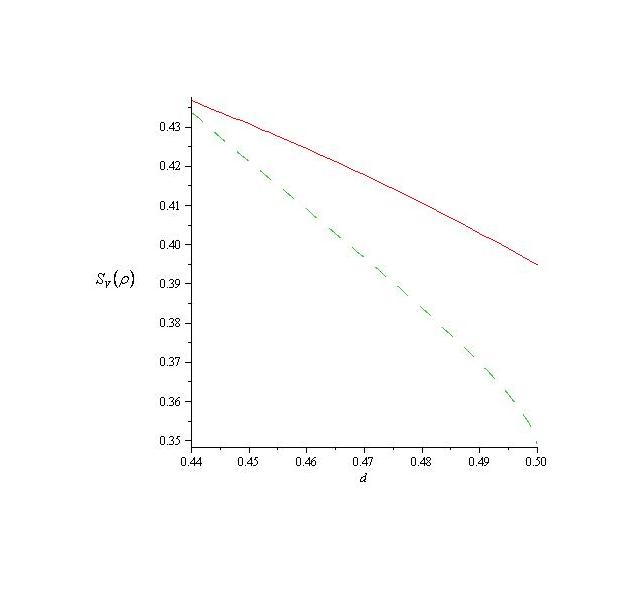}}
\caption{\footnotesize The ordinate of the figure represents the von - Neumann entropy of a state $\rho$, where solid line corresponds when $\rho = $ $\rho_{out}^{ab}$ and dotted line corresponds when $\rho = $ $\tau_{out}^{ab}$.} 
\end{figure}
From the above figure it is obvious that after distillation of the state $\rho_{out}^{ab}$, the filtered state $\tau_{out}^{ab}$ is less mixed than its original counterpart.
Mixedness, capacity of dense coding and fidelity of teleportation of the state $\tau_{out}^{ab}$ has already been calculated. In the following figure we plot these three quantities with respect to the parameter $d$.
\begin{figure}[!ht]
\centering
\resizebox{6cm}{6cm}{\includegraphics{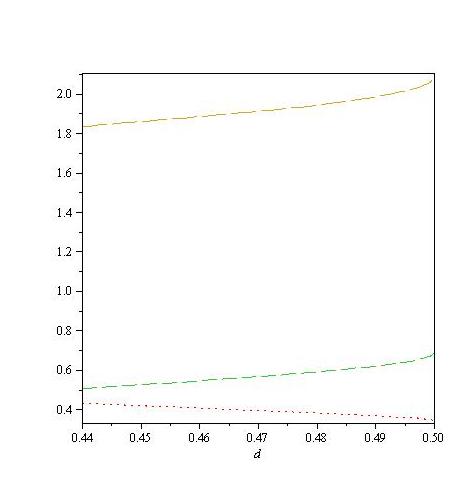}}
\caption{\footnotesize The dotted line represents mixedness of the filtered state, dashed line represents fidelity of teleportation and the long dashed line represents the capacity of dense coding of the said state.} 
\end{figure}\\\\
From figure it is clear that when mixedness of the state decreases both the capacities of dense coding and fidelities of teleportation of the states for different values of $d$, increases.
\section{Summary and Discussion:}
To summarize, We have studied the entanglement properties and the usefulness of a two qutrit output state
generated through Buzek - Hillery quantum cloning machine for quantum information process-
ing. Although we found that the optimal as well as non-optimal output states are not useful
for information processing protocols such as teleportation and dense coding, the distilled output
states using appropriate filters can be used as a entangled resource for information processing for certain range of machine parameters. It is interesting to note that while the distilled optimal
output state can only be used for teleportation, the distilled non-optimal output state can be
used for teleportation as well as dense coding. Surprisingly, the optimal teleportation fidelity
obtained using the distilled non-optimal output state exceeds the optimal teleportation fidelity
obtained using the distilled optimal output state for certain values of machine parameter $d$.

\end{document}